\documentclass[10pt,aps,prx,twocolumn,groupedaddress,longbibliography]{revtex4-2}
\usepackage{graphicx}
\usepackage{caption}
\DeclareCaptionLabelSeparator{dot}{. }
\makeatletter
\def\justified{
	\let\\\@normalcr
	\@rightskip\z@skip \rightskip\@rightskip
	\leftskip\z@skip
	\parindent 0em\relax
	\setlength{\parfillskip}{0pt plus 1fil}}
\DeclareCaptionJustification{justified}{\justified}
\usepackage{subcaption}
\usepackage{amsmath}
\usepackage{amssymb}
\usepackage{mathrsfs}
\usepackage{braket}
\usepackage{units}
\usepackage{ragged2e}
\usepackage[colorlinks,urlcolor=blue ,citecolor=blue ,linkcolor=blue ]{hyperref}
\usepackage{xcolor}
\usepackage{nicefrac}
\usepackage{nameref}
\usepackage{textcomp}
\usepackage{urwchancal}
\usepackage{float}
\usepackage{multirow}
\usepackage{tabularx}
\usepackage[normalem]{ulem}
\definecolor{darkgreen}{rgb}{0,0.5,0}

\newcommand{\permille}{\ensuremath{{}^\circ\mkern-5mu/\mkern-3mu_{\circ\circ}}}

\newcommand{\bs}{\boldsymbol}

\newcommand{\um}{\mu{\rm m}}

\newcommand{\us}{\mu{\rm s}}

\newcommand{\vB}{\ensuremath{\bs{B}}}
\newcommand{\vk}{\ensuremath{\bs{k}}}

\newcommand{\tr}{\ensuremath{t_{\rm r}}}
\newcommand{\tho}{\ensuremath{t_{\rm h}}}

\newcommand{\as}{\ensuremath{a_{s}}}
\newcommand{\add}{\ensuremath{a_{\rm dd}}}
\newcommand{\edd}{\ensuremath{\epsilon_{\rm dd}}}

\newcommand{\Dyb}{\ensuremath{^{164}}{\rm Dy}}

\newcolumntype{Y}{>{\centering\arraybackslash}X}

\begin{document}

\title{Competing triangular and stripe supersolid orders in a dipolar quantum gas}

\author{Karthik Chandrashekara$^{1}$, Christian G\"olzh\"auser$^{1}$, Lily Platt$^{1}$, Jianshun Gao$^{1}$, Julian~Kusch$^{1}$, Lennart Hoenen$^{1}$, Manon Ballu$^{1}$, Wyatt Kirkby$^{1}$, Lauriane Chomaz$^{1}$}	
\email[Contact author: ]{chomaz@uni-heidelberg.de}
\affiliation{%
    $^{1}$ Physikalisches Institut, Universit\"at Heidelberg, Im Neuenheimer Feld 226, 69120, Heidelberg, Germany. chomaz@uni-heidelberg.de}
        
\date{\today}
	
\begin{abstract}
Supersolids are exotic quantum states in which long-range phase coherence coexists, and may interplay, with emergent spatial orders. A particularly rich phase diagram featuring several competing spatial orders is predicted for dipolar supersolids with two-dimensional crystals, yet the experimental observation of this structural variety has remained limited. Here we experimentally form competing triangular and stripe density-modulated states in a quantum gas of highly magnetic atoms confined in a surfboard-shaped trap by tuning contact interaction strength and dipole orientation. We define a structural order parameter and study its statistical behavior. Thereby, we identify both the triangular and stripe phases and the transition between them, the associated critical behavior being marked by enhanced non-Gaussian fluctuations. Furthermore, we observe each spatial structure in both the phase-coherent supersolid regime and the phase-incoherent insulating one, near and far from the unmodulated-to-modulated transition, respectively.  Our results establish a versatile platform in which multiple phases of the two-dimensional-supersolid phase diagram, and more generally, intertwined symmetry-breaking phenomena, can be investigated.
\end{abstract}

\maketitle


\section{Introduction}

The emergence, competition, and coexistence of multiple broken symmetries underlie some of the richest phenomena in many-body physics, from condensed to active matter~\cite{Fradkin2015cto, Bowick2022stt}, yet the underlying mechanisms are challenging to isolate. Supersolids, which combine superfluid and crystalline order within a single quantum phase~\cite{Gross1957Unified,Gross1958Classical,Boninsegni2012Colloquium}, provide a pristine setting to address this challenge. 
First envisioned in the context of helium~\cite{Boninsegni2012Colloquium}, they have been realized on several platforms~\cite{Leonard2017Supersolid,li2017stripe,Bersano2019Experimental,Tanzi2019Observation,Boettcher2019Transient,Chomaz2019LongLived,Chomaz2022Dipolar,Chomaz2026Quantum,Putra2020spatial,Chisholm2026,Liebster2025supersolid,Trypogeorgos2025esi,Nigro2025sop, Xiang2024gme,Conti2023Chester}, most prominently in dipolar quantum Bose gases, where they are studied with an unprecedented level of control~\cite{Tanzi2019Observation,Boettcher2019Transient,Chomaz2019LongLived,Chomaz2022Dipolar,Chomaz2026Quantum}. Dipolar supersolids were first observed with one-dimensional (1D) crystal structures using cigar-shaped traps~\cite{Tanzi2019Observation,Boettcher2019Transient,Chomaz2019LongLived,Chomaz2022Dipolar,Chomaz2026Quantum}, and later on with two-dimensional (2D) crystals, specifically triangular arrays, using surfboard- and pancake-shaped traps~\cite{Norcia2021Twodimensional,Bland2022TwoDimensional,Chomaz2022Dipolar,Chomaz2026Quantum}.

Unlike their 1D counterparts, 2D dipolar supersolids are predicted to support multiple crystalline arrangements. These range from triangular arrays at low densities~\cite{Lu2015, Zhang2019Supersolidity, Hertkorn2021Pattern, Norcia2021Twodimensional,Bland2022TwoDimensional,Ripley2023Twodimensional}, to stripe or labyrinthine states at intermediate densities~\cite{Lu2015, Hertkorn2021Pattern, Ripley2023Twodimensional}, and honeycomb or pumpkin states at high densities~\cite{Lu2015, Zhang2019Supersolidity, Hertkorn2021Pattern, Ripley2023Twodimensional}. 
The various arrangements are expected to differ in their phononic and hydrodynamic responses, reflecting different manifestations of supersolidity~\cite{Ripley2023Twodimensional,Poli2024eoa,Blakie2025Dirac,SenarathYapa2025Anomalous}. The competition between arrangements gives rise to structural phase transitions~\cite{Seul1995, Li2016, Li2021, GiovannaReview} within the supersolid regime~\cite{Lu2015, Zhang2019Supersolidity, Ripley2023Twodimensional}.  Experimentally accessing these phases and the transitions connecting them provides a window into how global phase coherence interplays with rearranging spatial order, making it a major outstanding goal.

\begin{figure}[ht!]
    \centering
    \includegraphics[width=1.0\linewidth]{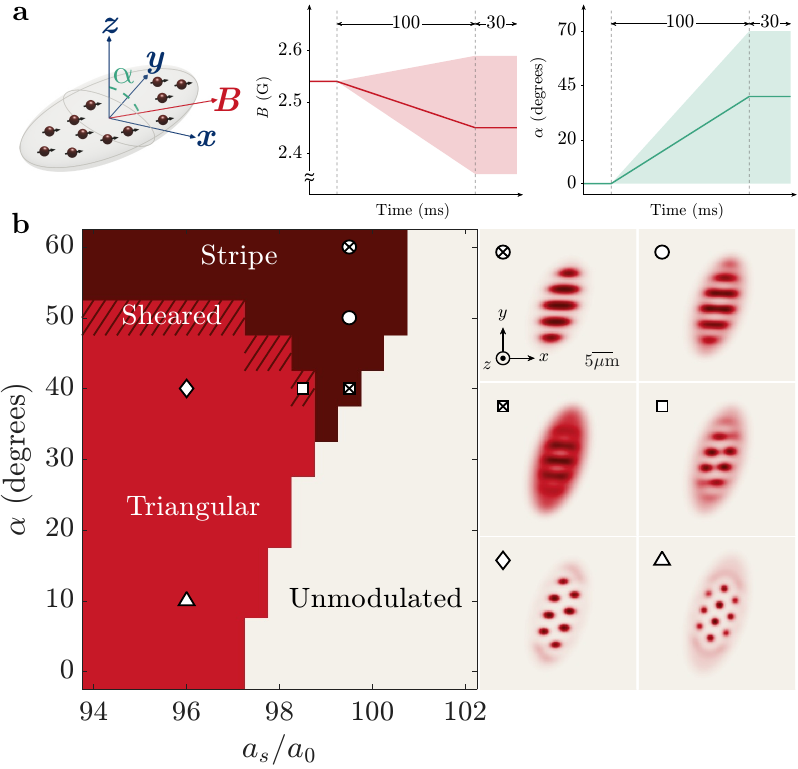}
    \caption{\textbf{Protocol and Phase Diagram} (a) Sketch of the gas geometry (left panel). The gas is confined in a shallow surfboard trap with tighter confinement along $z$. A uniform magnetic field is applied in the $xz$ plane with tunable magnitude $B$ and tilt angle $\alpha$. In the experiment, we ramp $B$ and $\alpha$ in 100\,ms, hold for 30\,ms (right panels).  (b) Phase diagram as a function of $a_s$ and $\alpha$ constructed from  TeGPE simulations with experimental settings. Alongside exemplary ground-state density profiles, cut at $z=0$ (parameters identified by symbols). Unmodulated, triangular, and stripe states are identified from such profiles (App.\,\ref{sec:simulations}).} 
    \label{fig:protocol}
\end{figure}

Previous experiments have almost exclusively considered dipoles aligned with the tight confinement axis~\cite{Norcia2021Twodimensional,Bland2022TwoDimensional,Tanzi2019Observation,Boettcher2019Transient,Chomaz2019LongLived,Chomaz2022Dipolar,Chomaz2026Quantum,Biagioni2022Dimensional}, see however~\cite{Wenzel2017Striped,He2025Observation}. Recent theory predicts that tilting them toward a weakly confined direction reshapes the phase diagram~\cite{Lima2025Supersolid}. It brings competing crystalline orders into the experimentally accessible density regime and alters the character of both the unmodulated-to-modulated and the structural transitions.

Here, we experimentally explore the rich landscape of states that spontaneously emerge in dipolar quantum gases confined in a surfboard-shaped trap upon slowly varying both the contact interaction strength and the dipole tilt angle, $\alpha$, relative to the tight confinement axis [Fig.\,\ref{fig:protocol}(a), Sec.\,\ref{sec:system}]. We observe density-modulated states with distinct 2D spatial order, changing from triangular arrays to stripes with increasing dipole tilt. By defining a structural order parameter, we identify the two ordered phases as well as a critical regime at intermediate tilts ($\alpha \approx 30-45^\circ$), associated with the structural transition between them~\cite{Chomaz2003, Pichon2006, Fink2017, Biagioni2022Dimensional, Beaulieu2025, Allemand2026, Chalopin2026, Beregi2026} (Sec.\,\ref{sec:structural_orders}). Close to the onset of density modulation, both the triangular and stripe phases display global phase coherence, establishing them as supersolids with distinct spatial order (Sec.\,\ref{sec:global_phase_coherence}). These observations are supported by theoretical predictions using an extended mean-field framework including quantum and thermal fluctuation effects (Sec.\,\ref{sec:system}), in qualitative agreement with the experimental behaviors.

\section{System and predictions}~\label{sec:system}

For this work, we use a degenerate quantum gas comprising
$1.4(1)\times10^{5}$ $^{164}\mathrm{Dy}$ atoms, confined in an anisotropic harmonic trap of in-plane and axial ($z$) frequencies $[39(1), 20(2)]\,$Hz and $ 97(1)\,$Hz, respectively (App.\,\ref{sec:preparation}).  A homogeneous magnetic field $\vB$  of magnitude $B=2.54\,$G oriented along $z$ ensures that the gas is initially in the unmodulated superfluid phase with a condensed fraction and temperature estimated to $64(3)\%$ and $62(4)$\,nK, respectively. 

The interatomic interactions can be dynamically controlled through the orientation and the magnitude of $\vB$ [Fig.\,\ref{fig:protocol}(a)]. The orientation of $\vB$, parameterized by the angle $\alpha$ between $\vB$ and the $z$ axis, sets the orientation of the atomic dipoles. The field magnitude $B$ sets the contact interaction strength, characterized by the scattering length $a_s$, through low-field Feshbach resonances~\cite{Chomaz2019LongLived,Chomaz2022Dipolar}. The in-plane component of $\vB$ defines the $x$ axis, which is rotated by $\approx16^\circ$ with respect to the trap axis (App.\,\ref{sec:control}). The dipolar interaction strength is fixed by the dipolar length $\add=131\,a_0$ of $\Dyb$. Hereafter, $\alpha$ is varied from $0^\circ$ to $70^\circ$, while $B$ is tuned from $2.36\,$G to $2.59\,$G, corresponding to an estimated variation of $a_s$ from $89\,a_0$ to $109\,a_0$ (App.\,\ref{sec:control}).


\begin{figure}[ht!] 
    \centering
    \includegraphics[width=1.0\linewidth]{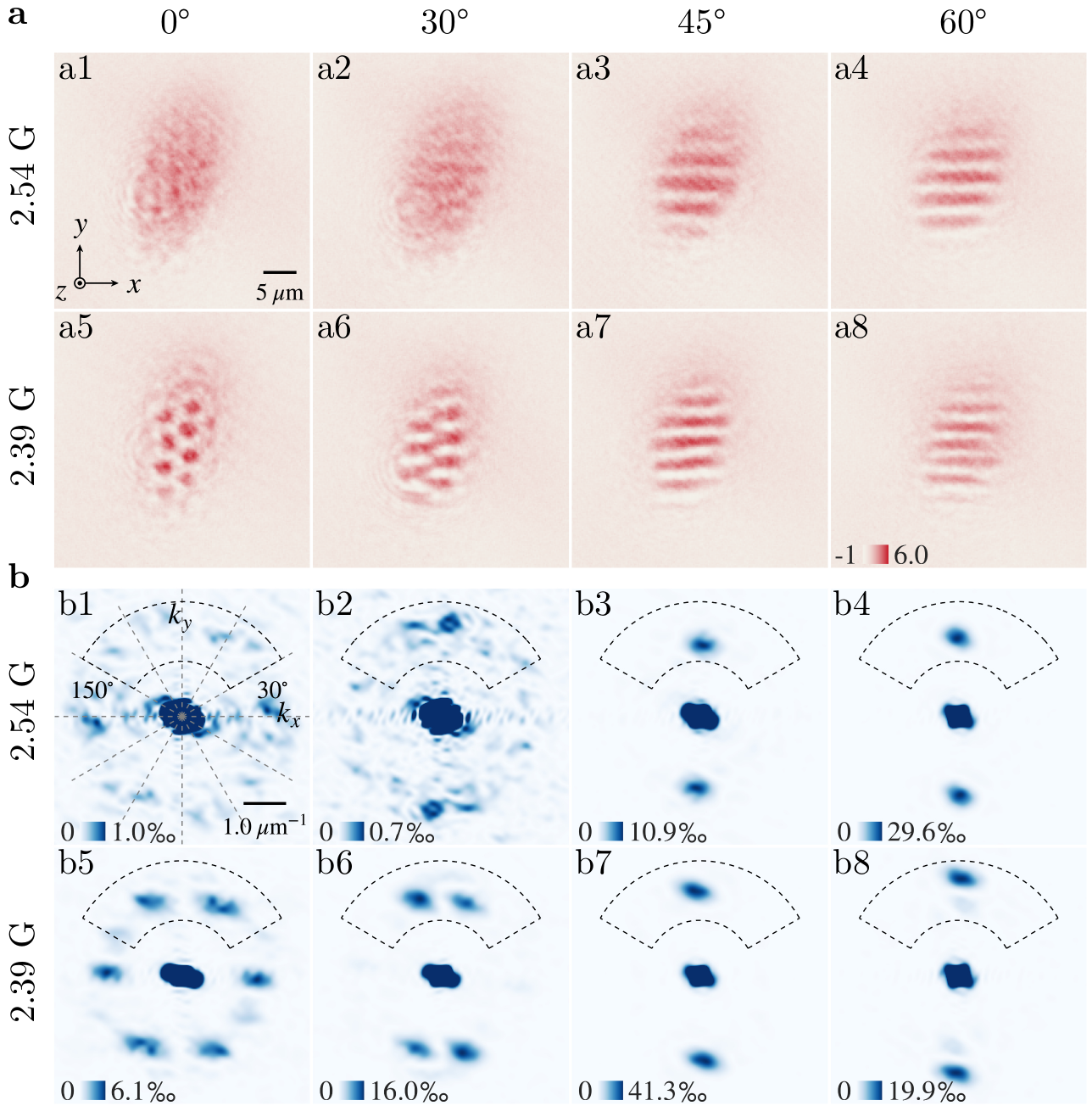}
    \caption{\textbf{State structure analysis} (a) Examples of experimental in-situ images after ramping to $B=2.54\,$G ($99.0\,a_0$, upper row) and $B=2.39\,$G ($90.8\,a_0$, lower row) and $\alpha =0^\circ,30^\circ,45^\circ,60^\circ$ from left to right. (b) Power spectra of the states shown in (a). $S(\vk)$ is normalized to the peak at $\vk=0$; the color scale is clipped to the maximum of the finite-$\vk$ peaks in the range $k_\rho \in [1.3, 2.7]\,\um^{-1}$ for visibility.}
    \label{fig:geom}    
\end{figure}

To estimate the equilibrium states expected under our experimental conditions as $\alpha$ and $a_s$ are varied, and to guide our measurements, we perform mean-field ground-state calculations, including the effects of thermal and quantum fluctuations, using a temperature-dependent extended Gross-Pitaevskii equation (TeGPE)~\cite{
SanchezBaena2023Heating,SanchezBaena2024SuperfluidSupersolid,Kusch2026} (App.\,\ref{sec:TeGPE} and~\ref{sec:simulations}). Figure~\ref{fig:protocol}(b) shows the calculated phase diagram as a function of $\alpha$ and $a_s$, together with exemplary in-plane density profiles (cut at $z=0$).

At large $a_s$, the ground state is unmodulated. Decreasing $a_s$ yields density-modulated states for which we identify three distinct regimes in $\alpha$. For small $\alpha < \alpha_1^{\rm (th)} \approx  35^\circ$, the modulated states consistently form almost equilateral triangular arrays of droplets (see $\triangle$), as previously observed in Refs.\,\cite{Norcia2021Twodimensional,Bland2022TwoDimensional}. 
For large $\alpha>\alpha_2^{\rm (th)} \approx 55^\circ$, stripe patterns, aligned along $x$, systematically form  (see $\otimes$). 
The values of $\alpha_1^{\rm (th)}$, $\alpha_2^{\rm (th)}$ cannot be readily inferred from uniform quasi-2D energetic considerations~\cite{Bourgeois2024eot,Baranov2008Theoretical,Baillie2015General,Blakieprivate}.

At intermediate angles $\alpha_1^{\rm (th)} \leq \alpha \leq \alpha_2^{\rm (th)}$, different structures spontaneously arise depending on $a_s$.
For $\as$ just below the unmodulated-to-modulated transition, stripes form (see $\boxtimes$). Upon decreasing $a_s$, an increasingly pronounced density modulation develops within each stripe (see $\bigcirc$), continuously transforming them into sheared triangular droplet arrays (see $\square$). In these sheared arrays, the droplets are aligned along $x$ and the triangle-shear angle appears to result from the interplay between the trap and dipole-tilt axes. 
Upon decreasing $a_s$ further, the arrays transform into more equilateral triangular arrangements, staggered along $y$ (see $\Diamond$).

Guided by these theoretical predictions, we experimentally explore the phase diagram by simultaneously ramping $a_s$ and $\alpha$ in $100\,$ms starting from our unmodulated quantum gas, and holding in trap for $30\,$ms [Fig.\,\ref{fig:protocol}(a)]. We then image the column density distribution $n(x,y)$ of the state via absorption imaging along the $z$-axis, either just after switching off the trap (in-situ measurements) or after $50\,$ms of free evolution [time-of-flight (TOF) measurements] (App.\,\ref{sec:imaging}). In Sec.\,\ref{sec:structural_orders}, we investigate the emergence of triangular and stripe structural orders by analyzing in-situ measurements. In Sec.\,\ref{sec:global_phase_coherence}, we probe the global phase coherence of these states using TOF measurements.

\section{Identifying structural orders}\label{sec:structural_orders}

Figure \ref{fig:geom}(a) shows exemplary in-situ images across the phase diagram. 
For large $B$ and small $\alpha$, we obtain mostly structureless density distributions, identified as unmodulated superfluid states (a1-a2). 
Although weak density fluctuations are present, they are predominantly random and only occasionally display local tendencies toward crystallization [see e.g.\,(a2)], see also Refs.\,\cite{Schmidt2021Roton,Hertkorn2021Density} and App.\,\ref{sec:insituanalysis}. Upon increasing $\alpha$ at fixed $B$, marked structures spontaneously appear in the density, here with stripe-like patterns along $x$ (a3,a4). Likewise, structures form upon decreasing $B$, now with different arrangements depending on $\alpha$, from triangular droplet arrays (a5,a6) to stripe-like modulations (a7,a8). In (a6), the droplets appear to stretch in the direction of the tilt. This effect stems from the predicted elongation of the droplets along $\vB$ combined with the axial integration of our imaging system. While the staggered droplet arrays remain distinguishable, this apparent droplet stretching with $\alpha$ prevents us from clearly distinguishing between sheared droplet arrays and stripes at large $\alpha$ in experimental images. Despite this caveat, our observations qualitatively follow the theoretical predictions of Fig.\,\ref{fig:protocol}(b), and give first hints of the formation of triangular droplet arrays and stripes in the experiment.

\begin{figure*}[ht!]
    \centering
    \includegraphics[width=0.80\linewidth]{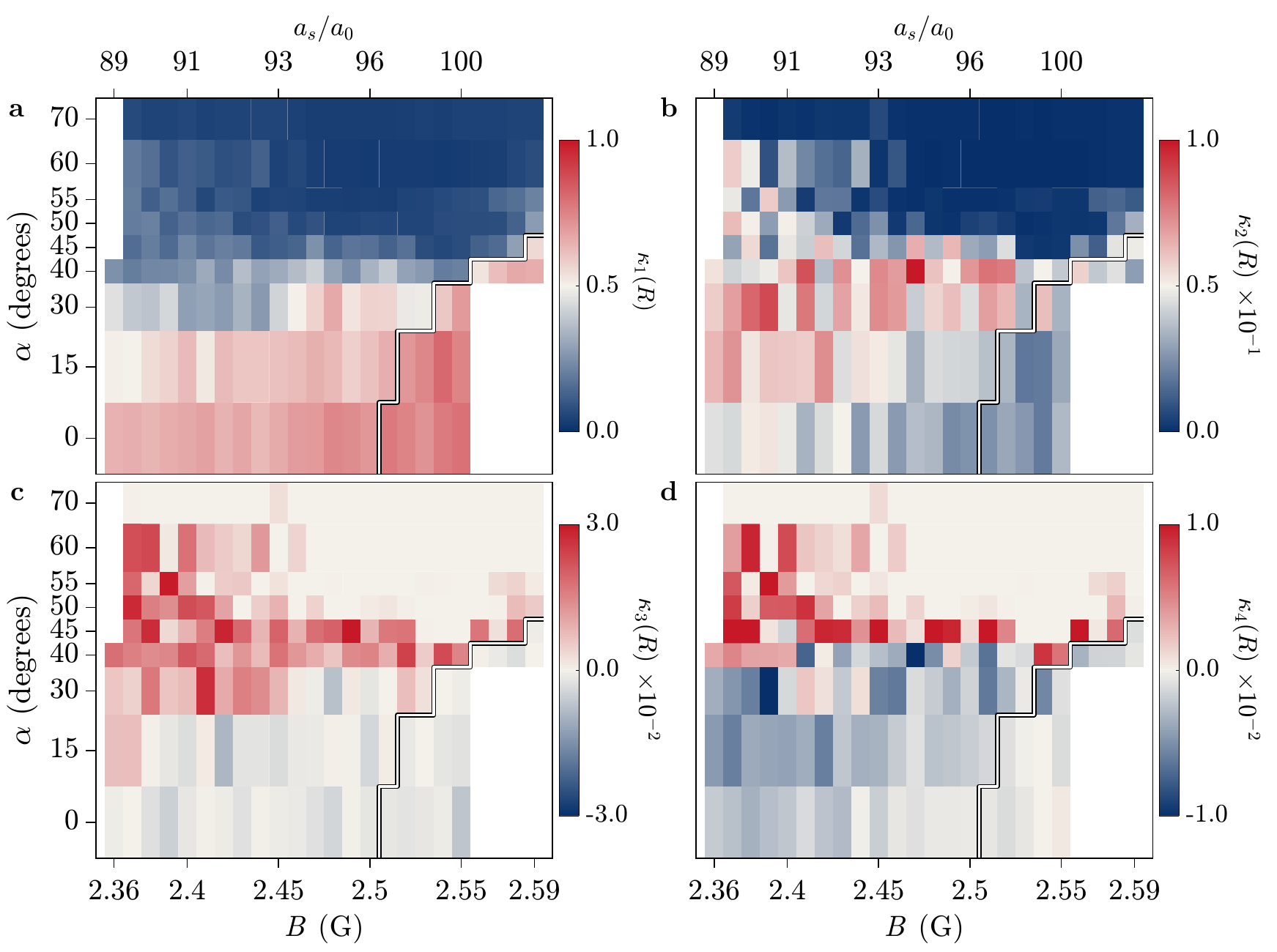}
    \caption{\textbf{Statistics of the structural order parameter} The $n^{\rm th}$-order cumulants $\kappa_n$ of the order parameter $R$ extracted from the power spectra of 30 independent experimental shots for varying $B$ and $\alpha$ with  (a) $n=1$ (mean), (b) $n=2$ (variance), (c) $n=3$ and (d) $n=4$ cumulants. The solid white line highlights the boundary between modulated and unmodulated states extracted from the S(\vk)-amplitude analysis (App.\,\ref{sec:insituanalysis}).}
    \label{fig:opstatistics}
\end{figure*}

To quantitatively analyze the structure in the individual density profiles, we compute their power spectrum $S(\vk) = S(k_x,k_y)= S(k_\rho,\theta) = \left| \mathcal{F}\{n(x,y) \} \right|^{2}$ where $\mathcal{F}$ is the 2D Fourier transform, $\vk$ the in-plane wavevector, $k_x$ and $k_y$ ($k_\rho$ and $\theta$) its Cartesian (polar) coordinates. Note that for a translation-invariant system, $S(\vk)$ also corresponds to the 2D Fourier transform of the density-density correlation function~\cite{Pitaevskii16}. 
Figure \ref{fig:geom}(b) shows the power spectra of the in-situ images of Fig.\,\ref{fig:geom}(a). While the peak at $\vk=0$ encodes the overall shape of the cloud, density structures yield peaks at finite $\vk$. 
The amplitudes of the latter peaks provide a measure of the modulation strength, see the color scales of (b3-b8) in contrast to (b1-b2), and are used below to extract the modulation-onset boundary (Fig.\,\ref{fig:opstatistics} and~\ref{fig:phasecoherence}; App.\,\ref{sec:insituanalysis}).

Beyond their amplitude, the number and geometric arrangement of the finite-$\vk$ peaks encode the structure of the density pattern and its regularity. In the absence of a clear density modulation [Fig.\,\ref{fig:geom}(b1-b2)], $S(\vk)$ shows weak finite-$\vk$ peaks with fluctuating number, arrangement, and amplitudes. Instead, each structural arrangement displays a characteristic spectral signature.
Stripe-like patterns [Fig.\,\ref{fig:geom}(b3,b4,b7,b8)] produce two dominant finite-$\vk$ peaks near $k_x= 0$. 
Triangular arrays with axial magnetization [$\alpha=0$, Fig.\,\ref{fig:geom}(b5)] yield six finite-$\vk$ peaks of similar amplitude and almost regular $60^\circ$ spacing, with two of these peaks near $k_y=0$.  Staggered triangular arrays with finite $\alpha$ [Fig.\,\ref{fig:geom}(b6)] yield four finite-$\vk$ peaks again of similar amplitude, grouped into closely spaced pairs near $k_x=0$. 
The angular spacing of the peaks depends primarily on the apparent stretching of the individual droplets in $n(x,y)$ (and thus on $\alpha$). We note that excitations of these regular structures, as observed in some shots, result in unequal amplitudes of the finite-$\vk$ peaks or the emergence of additional peaks (App.\,\ref{sec:furtherinsituexamples}).


Motivated by these characteristic spectral signatures, we identify an order parameter $R$ for the triangular-to-stripe structural transition. We extract the amplitudes, $S_1$ and $S_2$, of the two dominant finite-$\vk$ peaks in the upper sector of $S(k_x,k_y>0)$, specifically in the range $\theta \in [30^\circ,150^\circ]$ and $k_\rho \in [1.3, 2.7]\,\um^{-1}$ [dashed lines in Fig.\,\ref{fig:geom}(b)], and define $R=S_2/S_1$. Stripe-like patterns yield $R\approx 0$, at the percent level in our data, whereas triangular arrangements produce substantially larger values. Perfect triangular arrangements have $R = 1$ while excitations yield reduced $R$ values, typically to the several tens of percent. The axial integration of our imaging further increases the sensitivity of $R$ to excitations with increasing $\alpha$ (see e.g.\,App.\,\ref{sec:furtherinsituexamples},~\ref{sec:experimentalRDistributions} and \ref{sec:theoryRDistributions}). We note that the  order parameter for the triangular-to-stripe transition under dipole tilt remains theoretically unknown, as no Landau theory framework has yet been developed~\cite{Biagioni2022Dimensional,Allemand2026,Blakieprivate}.

In finite-size systems, the statistical properties of an order parameter, in particular its higher-order cumulants, contain signatures of critical behavior~\cite{Allemand2026, Chalopin2026, Chomaz2003, Pichon2006,  Fink2017, Beaulieu2025, Biagioni2022Dimensional, Beregi2026}. We therefore analyze the first four cumulants of $R$ across the parameter space from 30 independent realizations per parameter set (Fig.\,\ref{fig:opstatistics}). The mean of $R$ [Fig.\,\ref{fig:opstatistics}(a)] evolves from $\approx 1$ to  $\approx 0$ upon tilting the dipoles,  indicating a transition from triangular to stripe arrangements. For $\alpha \approx 30-45^\circ$, $R$ takes intermediate mean values and shows enhanced [increased variance; Fig.\,\ref{fig:opstatistics}(b)] and non-Gaussian [non-zero higher-order cumulants;  Fig.\,\ref{fig:opstatistics}(c,d)] fluctuations.  These features are characteristic of critical behavior in finite systems and indicate competition between two orders through low-energy metastable states or soft modes depending on whether the transition is first or second order~\cite{Allemand2026, Chalopin2026, Chomaz2003, Pichon2006,  Fink2017, Beaulieu2025, Biagioni2022Dimensional, Beregi2026}.

Outside this intermediate regime, fluctuations are reduced and higher-order cumulants approach zero [Fig.\,\ref{fig:opstatistics}(b,c,d)], indicating well-defined orders. Although our imaging cannot directly distinguish between sheared triangular states and stripes (see above), the reduced fluctuations of $R$, together with our theory [Fig.\,\ref{fig:protocol}(b)], indicate the emergence of stripe order at large $\alpha$ (see also App.\,\ref{sec:experimentalRDistributions} and ~\ref{sec:theoryRDistributions}). This reduced-fluctuation signature does not allow us to precisely pinpoint the onset angles $\alpha_{1,2}^{\rm (exp)}$ in experiment but places upper bounds on them. We identify the stripe phase for $\alpha \geq 45^\circ \geq \alpha_{1}^{\rm (exp)}$, in a range of $a_s$ just below the unmodulated-to-modulated transition. Its $a_s$ range widens as $\alpha$ increases, and dominates for $\alpha \geq 70^\circ \geq \alpha_{2}^{\rm (exp)}$. The extracted bounds are consistent with our theory predictions (see Sec.\,\ref{sec:system}). 

Our statistical analysis establishes the existence of density-modulated states with both triangular and stripe order, as well as a phase transition connecting these competing arrangements. The underlying phase diagram is rich, involving intertwined unmodulated-to-modulated, triangular-to-stripe, and, as shown below, phase-coherent-to-incoherent transitions. Because our protocol concomitantly ramps $\alpha$ and $a_s$, the crossings of these transitions are coupled, complicating the interpretation of the observed critical behavior. Furthermore, the use of a fixed ramp time implies varying ramp rates, also affecting the fluctuation behavior.
These findings motivate future studies toward a comprehensive characterization of the transitions and their criticality.

\section{Probing global phase coherence}\label{sec:global_phase_coherence}
\begin{figure*}[ht!] 
    \centering
    \includegraphics[width=1.0\linewidth]{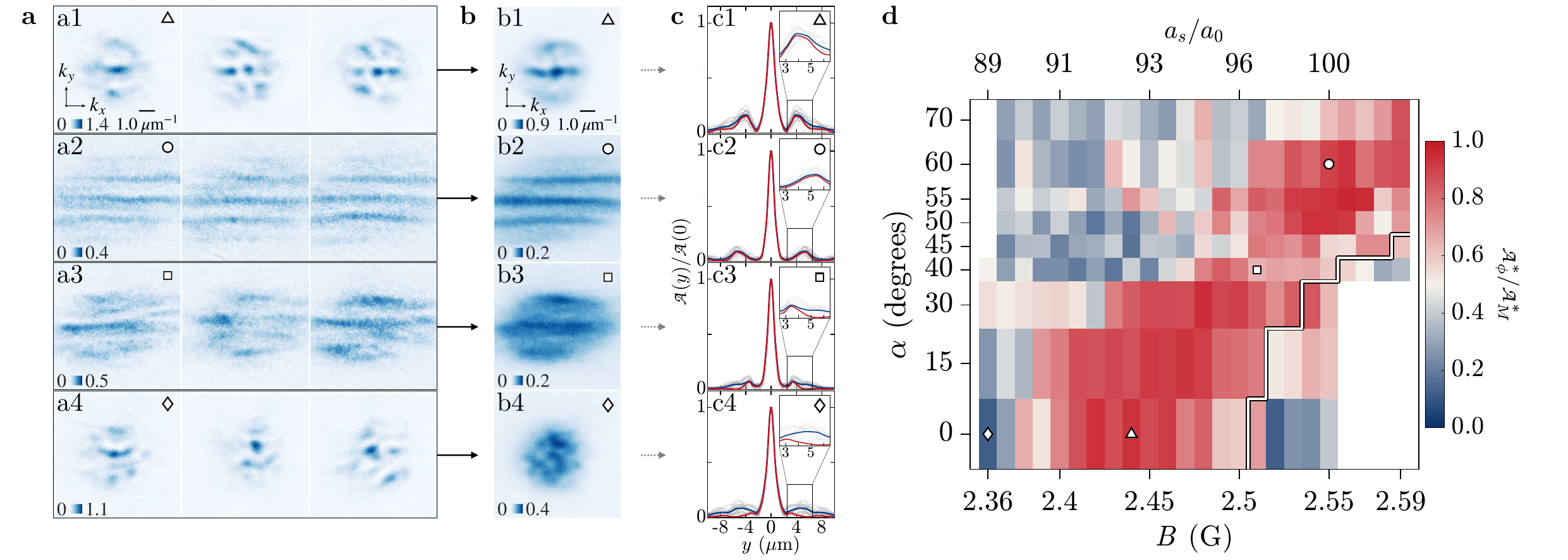}
     \caption{\textbf{Global phase coherence} (a) Examples of single TOF density profiles, with (b) corresponding averaged images over 20 shots. The TOF images are shown as a function of the in-plane momentum components, $k_x$, $k_y$, using the free-expansion formula (App.\,\ref{sec:coherence}).  (c) Corresponding 1D Fourier transforms of the individual density profiles integrated along $k_x$ (grey lines) together with $\mathcal{A}_\phi(y)$ (red line), $\mathcal{A}_M(y)$ (blue line). The inset in (c) enlarges the range around the $y>0$ side peak.  Figures are shown for: $B = 2.44\,\mathrm{G}$, $\alpha = 0^\circ$ (1,$\triangle$), $B = 2.55\,\mathrm{G}$, $\alpha = 60^\circ$ (2,$\bigcirc$), $B = 2.51\,\mathrm{G}$, $\alpha =40^\circ$ (3,$\square$), $B = 2.36\,\mathrm{G}$, $\alpha = 0^\circ$ (4,$\Diamond$). (d) $\mathcal{A}_\phi^*/\mathcal{A}_M^*$ as a measure of global phase coherence over different $B$ and $\alpha$. The solid white line highlights the boundary between modulated and unmodulated states from the in-situ analysis (App.\,\ref{sec:insituanalysis}).}
     \label{fig:phasecoherence}
 \end{figure*}

 We now examine the global phase coherence of the diverse density-modulated states realized in our experiment, a key signature of their supersolid character~\cite{Chomaz2019LongLived,Tanzi2019Observation,Boettcher2019Transient} (Fig.\,\ref{fig:phasecoherence}). We image the gas after 50\,ms  TOF, during which the different density-modulated regions interfere (App.\,\ref{sec:sequence}). The global phase coherence is typically revealed through the reproducibility of the resulting interference patterns between shots ~\cite{Ilzhofer2021pci, Chomaz2019LongLived, Tanzi2019Observation, Boettcher2019Transient, Hadzibabic2004Interference}. Yet, the underlying in-situ density arrangement, which varies in our case, also affects the interference geometry. Therefore, fluctuations in both the phase relation and density arrangement impact the reproducibility of the TOF patterns.


 We distinguish four characteristic regimes from individual TOF images and their 20-shot averages [Fig.\,\ref{fig:phasecoherence}(a,b)]. Near the unmodulated-to-modulated transition, both triangular ($\triangle$,a1,b1) and stripe ($\bigcirc$,a2,b2) states exhibit reproducible interference patterns that remain visible upon averaging, establishing their global phase coherence. The distinct interference geometries reflect the underlying in-situ arrangements: triangular states yield a central peak at $\vk=0$ surrounded by finite-$\vk$ peaks yielding a modulated annulus on average, whereas stripe states yield three main streaks along $k_y$. By contrast, in the critical regime ($\square$,a3,b3), the interference patterns fluctuate from shot to shot. A residual structure survives averaging (b3), albeit with reduced contrast and less-defined features than in (b1,b2). We attribute this behavior to the strong fluctuations of the in-situ arrangements revealed by the structural analysis [Fig.\,\ref{fig:opstatistics}], rather than to a loss of phase coherence. Finally, farther from the modulation onset ($\Diamond$,a4,b4), the interference patterns become more complex and irreproducible. They disappear upon averaging, indicating the loss of global phase coherence and the emergence of an insulating state (see also App.\,\ref{sec:furtherphasecoherence}).

To quantify the degree of phase coherence, we perform a Fourier analysis of the TOF patterns [Fig.\,\ref{fig:phasecoherence}(c)].
For each shot $i$, we integrate the TOF image along $k_x$ and compute the 1D Fourier transform, $\mathcal{A}^{(i)}(y)$, of the resulting profile. The amplitude and phase of $\mathcal{A}^{(i)}(y)$ at the finite-$y$ peaks of $|\mathcal{A}^{(i)}(y)|$ [Fig.\,\ref{fig:phasecoherence}(c), grey lines] encode the interference patterns along $k_y$. By performing averages over all shots $\langle ...\rangle_i$, we extract two complementary observables: the modulation amplitude, $\mathcal{A}_M(y)=\langle |\mathcal{A}^{(i)}(y)|\rangle_i$, which quantifies the presence of interference fringes in the individual TOF images, and the phase amplitude $\mathcal{A}_\phi(y)=|\langle \mathcal{A}^{(i)}(y)\rangle_i|$, which probes their shot-to-shot phase stability [Fig.\,\ref{fig:phasecoherence}(c) blue and red lines]~\cite{Chomaz2019LongLived,Tanzi2019Observation,Boettcher2019Transient,Ilzhofer2021pci}. More precisely, these properties are quantified by the values of $\mathcal{A}_M(y)$ and $\mathcal{A}_\phi(y)$ at the finite-$y$ peaks defined as $\mathcal{A}_M^*$ and $\mathcal{A}_\phi^*$, respectively (App.\,\ref{sec:coherence}). 
Note that $\mathcal{A}_\phi(y)$ also coincides with the norm of the 1D Fourier transform of the averaged TOF images [Fig.\,\ref{fig:phasecoherence}(b)] integrated along $k_x$.

This analysis quantitatively supports the observations of Fig.\,\ref{fig:phasecoherence}(a,b) [see also App.\,\ref{sec:furtherphasecoherence}]. For phase-coherent triangular (c1) and stripe (c2) states, $\mathcal{A}_M$ and $\mathcal{A}_\phi$ nearly coincide and show pronounced finite-$y$ peaks, yielding $\mathcal{A}_M^* \approx \mathcal{A}_\phi^*$ (inset). In the critical case (c3), $\mathcal{A}_M$  shows a broad, low-amplitude bump at finite $y$, whereas $\mathcal{A}_\phi$ is confined to small $y$, resulting in $\mathcal{A}_M^* > \mathcal{A}_\phi^*$ (inset). The phase-incoherent case (c4) follows a similar trend, with a further suppression of $\mathcal{A}_\phi$, and $\mathcal{A}_M^* \gg \mathcal{A}_\phi^*$ (inset). 

Although it is affected by structural fluctuations, we use $\mathcal{A}_\phi^*/\mathcal{A}_M^*$ as an indicator for global phase coherence~\cite{Chomaz2019LongLived,Ilzhofer2021pci} and track its evolution across the parameter space  [Fig.\,\ref{fig:phasecoherence}(d)]. For most tilt angles, except $\alpha=40-45^\circ$, $\mathcal{A}_\phi^*/\mathcal{A}_M^*$ remains large ($\approx 0.7-1$) over a range of $\as$ below the modulation-onset threshold (white line). Combined with our structural analysis [Fig.\,\ref{fig:opstatistics}], this identifies phase-coherent triangular arrays ($\alpha=0-30^\circ$), and phase-coherent stripes ($\alpha=50^\circ-70^\circ$). The coherent region extends over $\gtrsim 5\,a_0$ in $a_s$ before $\mathcal{A}_\phi^*/\mathcal{A}_M^*$ decreases, signaling a loss of global phase coherence for both structural arrangements, as expected from the predicted increasing density-modulation contrast [Fig.\,\ref{fig:protocol}(b)]~\cite{Chomaz2019LongLived,Ilzhofer2021pci,Wenzel2017Striped,Baillie2018Droplet,Roccuzzo2019Supersolid,Chomaz2026Quantum,Chomaz2022Dipolar,Tanzi2019Observation,Boettcher2019Transient}.

Near the unmodulated-to-modulated transition, $\mathcal{A}_\phi^*/\mathcal{A}_M^*$ also decreases. We attribute this behavior to the weaker density modulation in this regime, which reduces both $\mathcal{A}_M^*$ and $\mathcal{A}_\phi^*$, rendering their ratio particularly sensitive to noise (App.\,\ref{sec:coherence}). In the intermediate-tilt regime ($\alpha=40-45^\circ$), the strong fluctuations of the in-situ structures (Fig.\,\ref{fig:opstatistics}) likewise reduce $\mathcal{A}_M^*$ and $\mathcal{A}_\phi^*$ [Fig.\,\ref{fig:phasecoherence}(c3), App.\,\ref{sec:coherence}], and complicate the quantitative interpretation of $\mathcal{A}_\phi^*/\mathcal{A}_M^*$ in this regime.
Note that the TOF interference patterns contain considerably richer spatial and statistical information than captured by the average coherence indicator considered here. In particular, a statistical analysis may provide complementary information on the phase-coherent-to-incoherent transition and its interplay with the structural transition.

\section{Conclusions and outlook}

In conclusion, we realized both triangular and stripe density-modulated states in a dipolar quantum gas by tuning the contact interaction strength and dipole tilt. We identified each spatial order through a statistical analysis of a structural order parameter. We further established their insulating or supersolid nature by measuring global phase coherence. These states offer a promising platform to investigate how superfluid properties intertwine with crystalline order. In particular, the differently arranged supersolids are expected to host distinct collective excitations and hydrodynamic responses, providing access to this interplay, 
and a further probe of the crystalline arrangement~\cite{Poli2024eoa,Blakie2025Dirac,SenarathYapa2025Anomalous,Cook2026Excitations,Norcia2022Can}.  
Dipole tilting also implies anisotropic superfluid behaviors, which could be revealed in collective excitation spectra or in vortex dynamics, offering interesting directions to explore experimentally~\cite{Natale2019Excitation,Guo2019Lowenergy,Tanzi2019Supersolid,Hertkorn2021Density,Roccuzzo2020Rotating,Casotti2024Observation,Poli2025Synchronization}.

Our system also offers a rich landscape of phase transitions. Triangular-to-stripe, unmodulated-to-modulated, and phase-coherent-to-incoherent transitions provide a unique opportunity to investigate the interplay between different symmetry-breaking mechanisms. Control over interaction strength, dipole tilt, trap geometry, atom number, and temperature enables exploring different regimes, where the transitions may be of either first- or second-order character~\cite{Lima2025Supersolid,SanchezBaena2023Heating,SanchezBaena2024SuperfluidSupersolid}. Larger statistical datasets and tailored ramp protocols will enable a quantitative determination of the transitions and their order. Such studies could reveal how structural and superfluid orders compete or coexist, providing deep insight into quantum fluids with intertwined orders.

    \begin{acknowledgments}
        The authors thank B. Blakie, D. Baillie, E. Poli, A. Recati, S. Stringari, D. Clément, M. Allemand, F. Kaufmes, R. Dolbeault, G. Su, and P. Lunt for helpful discussions and collaboration on related work, and in particular M. Allemand and P. Lunt for comments on the manuscript. They acknowledge support by the European Research Council (ERC) under the European Union’s Horizon Europe research and innovation program under grant number 101040688 (project 2DDip), by the Deutsche Forschungsgemeinschaft (DFG, German Research Foundation), through SFB 1225 ISOQUANT (Project-ID 273811115), the ANR-DFG  bilateral project DISQUTT (Project-ID 530000649) and under Germany's Excellence Strategy -- EXC 2181/1 -- 390900948 (the Heidelberg STRUCTURES Excellence Cluster), and by the state of Baden-W{\"u}rttemberg through bwHPC and the German Research Foundation (DFG) through grant INST 35/1597-1 FUGG. W.K. acknowledges the support of the Natural Sciences and Engineering Research Council of Canada (NSERC), [funding reference number PDF-577924-2023].  Views and opinions expressed are however those of the authors only and do not necessarily reflect those of the European Union or the European Research Council. Neither the European Union nor the granting authority can be held responsible for them.  
    \end{acknowledgments}

\section*{Data Availability}
The data presented in this study are available from the corresponding author upon reasonable request.

\appendix

\section{Preparation of the initial samples}\label{sec:preparation}

The experimental sequence starts with loading a three-dimensional (3D) magneto-optical trap (MOT) using 626\,nm light from a slow jet of dysprosium atoms prepared via a 2D-MOT and push-beam scheme as described in Ref.\,\cite{Jin2023Two}. 
The 3D-MOT is made using five $12~\mathrm{mm}$-waist beams~\cite{Ilzhofer2018Two}, with two retro-reflected horizontal beams and one vertical beam pointing upward, and a magnetic gradient that is created by an anti-Helmholtz pair of coils along $z$. The power and detuning for the horizontal and vertical beams are controlled independently. During  the $5\,\mathrm{s}$ loading stage, the horizontal (vertical) beams are operated at $\approx200$\,mW ($\approx11$\,mW) power and $-38~\mathrm{\Gamma_{626}}$ ($-36~\mathrm{\Gamma_{626}}$) detuning.  The gradient magnitude is $0.4\,$G/cm in $xy$. A vertical offset field of  $B=0.5\,$G is applied. 
Then, the MOT is compressed in 160\,ms: the power and detuning in all beams are decreased to less than 1\% of the initial power and to $-32\,\mathrm{\Gamma_{626}}$. The magnetic gradient is increased to $0.65\,$G/cm, while the offset is suppressed.

Next, the atoms are transferred into an optical dipole trap (ODT) beam at 1064\,nm wavelength with a waist of $30\,\mathrm{\mu m}$ (ODT1) propagating along $y'$. 
The beam pointing is modulated along $x'$ at 10\,kHz by modulating the RF frequency of an acousto-optical deflector. An imaging system converts the resulting angular deflection into a position modulation~\cite{Ahmadi2005Geoeffects}. This effectively increases the beam waist in $x'$, thus changing the beam aspect ratio and enhancing the initial trap volume.  The beam aspect ratio can be tuned from 1 to 9.4 through the tuning of the RF frequency range. The ODT beam is ramped up to full power (34\,W) and maximal modulation range within 5\,ms. The MOT light and gradient are abruptly switched off 70\,ms later. After another 15\,ms, a $z$-offset field of $B=2.54\,$G is switched on. We load $3.4(1)\times10^{7}$ atoms into the ODT in this manner.

Evaporative cooling then proceeds in three stages utilizing a crossed ODT formed by ODT1 and a second beam (ODT2) of a larger waist of $60\,\mathrm{\mu m}$ and maximal power of 9\,W.  First, the power and aspect ratio of ODT1 are ramped down while ODT2 power is ramped up, trapping the atoms in the crossing region of the two beams. Second, the power of both beams is reduced further. 
Third, the ODT1 aspect ratio is ramped back up, while the powers of both beams are modified. This creates the final surfboard trap with trap frequencies $[\nu_{x'},\nu_{y'},\nu_z] = [39(1), 20(2), 97(1)]$ Hz. 
We hold the cloud in this trap for 1.6\,s to allow for thermalization. The $z$ offset field is maintained during the full evaporation sequence.

The resulting degenerate quantum gas is the starting point of the experiment. It comprises a total of $1.4(1)\times10^{5}$ $^{164}\mathrm{Dy}$ atoms with an estimated $64(3)\%$ condensed fraction and temperature of $62(4)$\,nK. These values are obtained from a bimodal fit to the density distributions of the gas after a $20\,\mathrm{ms}$ TOF measured by resonant low-intensity absorption imaging on the 421\,nm transition along a horizontal axis using a single-lens ($f=100\,\mathrm{mm}$) setup providing a magnification of 2.2. 

\section{Control of the magnetic field and scattering length}\label{sec:control}

A uniform magnetic field is created at the position of the atoms by three pairs of coils - a pair of circular coils along $z$ and two pairs of rectangular coils in $x$ and $y$ directions.  
This allows us to freely tune the orientation of $\vB$ in space. Here, we limit the tunability of $\vB$ to the $xz$ plane and define $\alpha$ as the angle between $\vB$ and $z$. The angle of $\approx 16^\circ$  between the $x$ and $x'$ axes comes from the tilt between the ODT beams and the coil axis.

Each coil pair is calibrated via Feshbach loss spectroscopy on three resonances close to our working fields, at $1.306\,$G,  $2.591\,$G and $2.803\,$G~\cite{Chomaz2019LongLived}. Preliminary calibrations via RF spectroscopy were also conducted. Feshbach spectroscopy is performed on a thermal gas of $1.3\times10^{6}$ atoms at a few microkelvin. We measure atom losses after quenching $B$ to the target value with the coil pair under calibration and holding for up to $1.5\,$s.

We convert  $B$ to $a_s$ using
\begin{equation}
    a_s=a_\mathrm{bg}\prod_i \left(1-\frac{\Delta_i}{B-B_{0,i}}\right)
\end{equation}
with $a_\mathrm{bg}$ the background scattering length, $B_{0,i}$ and $\Delta_i$ the position and width of the resonance $i$. We include 14 resonances between $1\,$G and $7\,$G and 1 broad resonance at $77\,$G~\cite{Chomaz2019LongLived,Maier2015Emergence,Baumann2014Observation,Chomaz2022Dipolar}. We adjust $a_\mathrm{bg}$ to best match the unmodulated-to-modulated transition between TeGPE simulations and experiment for $\alpha \in [0^\circ,40^\circ]$. This yields $a_\mathrm{bg} = 73.6(9)\,a_0$, a value consistent with previous estimates within uncertainties~\cite{FerrierBarbut2018Scissors,Chomaz2019LongLived}.

Statistical and systematic uncertainties on the magnetic field produced by each coil pair are estimated to be $\lesssim 10\,$mG. These result in uncertainties on $B$ and $\alpha$ on the order of $\sim 10\,$mG and $\pm 1^\circ$. Accounting for these uncertainties, the field range $[2.36, 2.58]\,$G maps onto $a_s \in [89 \pm 1, 109\pm 4]\,a_0$. At  $2.59\,$G, $a_s$ is ill-determined because of the nearby resonance.

\section{Temperature-dependent extended Gross-Pitaevskii theory}\label{sec:TeGPE}
To perform numerical simulations, we model our system using a temperature-dependent extended Gross-Pitaevskii equation (TeGPE)~\cite{Waechtler2016Quantum, Chomaz2016QuantumFluctuationDriven,
SanchezBaena2023Heating,SanchezBaena2024SuperfluidSupersolid,He2025Accessing,Kusch2026},

\begin{equation}
    \mathrm{i}\hbar\frac{\partial \Psi(\textbf{r},t)}{\partial t}=\Biggl[H_0+H_{\mathrm{int}}+H_{\mathrm{Q}}+H_{\mathrm{T}}\Biggr]\Psi(\textbf{r},t)\label{eq:TeGPE}
\end{equation}

where $\hbar$ is the reduced Planck constant and 

\begin{align}
    H_0=&\;-\frac{\hbar^2\nabla^2}{2m}+V(\textbf{r})\label{eq:H0}\\
    H_{\mathrm{int}}=&\;\int\mathrm{d}\textbf{r}'U(\textbf{r}-\textbf{r}')|\Psi(\textbf{r}',t)|^2\label{eq:Hint}
\end{align}

are the single-particle and two-body terms, respectively, with $m$ the atomic mass and $V(\textbf{r})$ the (harmonic) trapping potential. The interaction potential is given by,

\begin{equation}
    U(\textbf{r}) = g\delta(\textbf{r})+\mu_0\mu_m^2\frac{1-3\cos^2\vartheta}{4\pi|\textbf{r}|^3}\label{eq:Uint},\;
\end{equation}

 with the contact parameter $g=4\pi\hbar^2a_s/m$, the magnetic permeability $\mu_0$, the magnetic moment of \Dyb { }$\mu_m=9.98\mu_B$, where $\mu_B$ is the Bohr magneton, and $\vartheta = 90^\circ - \alpha$ is the angle between the $z$-axis and \textbf{r}. The Fourier transform of the interaction potential is
\begin{equation}
    \tilde{U}(\textbf{k})= g+\frac{\mu_0\mu_m^2}{3}\left(3\cos^2\alpha_\textbf{k}-1\right)
\end{equation}

where $\alpha_\textbf{k}$ is the angle between $\textbf{k}$ and the dipole polarization axis. The beyond-mean-field quantum fluctuation contribution is
\begin{equation}
H_{\mathrm{Q}}=\gamma_{\mathrm{Q}}|\Psi(\textbf{r},t)|^3\label{eq:HLHY}\;,
\end{equation}

with $\gamma_{\mathrm{Q}}=\frac{32}{3}g\sqrt{a_s^3/\pi}(1+3\epsilon_{\mathrm{dd}}^2/2)$ \cite{Schuetzhold2006Meanfield,Lima2011Quantum,Lima2012Beyond}. The thermal fluctuation contribution, captured by $H_{\mathrm{T}}$, is calculated by considering the Bogoliubov quasiparticle occupation number in the grand-canonical ensemble \cite{Pitaevskii16} and by using a local-density approximation \cite{SanchezBaena2023Heating,SanchezBaena2024SuperfluidSupersolid,He2025Accessing}
\begin{equation}    H_{\mathrm{T}}=\int\frac{\mathrm{d}\textbf{k}}{(2\pi)^3}\frac{\tilde{U}(\textbf{k})}{(\mathrm{e}^{\beta \epsilon_\textbf{k}}-1)}\frac{\tau_\textbf{k}}{\epsilon_\textbf{k}}\label{eq:HTherm}\;,
\end{equation}

where the free-particle [Bogoliubov] dispersion is $\tau_\textbf{k}=\hbar^2k^2/(2m) \ \left[\epsilon_\textbf{k}(\textbf{r})=\sqrt{\tau_\textbf{k}(\tau_\textbf{k}+2\tilde{U}(\textbf{k})|\Psi(\textbf{r})|^2)} \right]$. To evaluate the integral in Eq.\,\eqref{eq:HTherm}, we use an ansatz that well-approximates the density dependence,

\begin{equation}
    H_{\mathrm{T}}\approx A[a_s,T]\,\mathrm{e}^{-B[a_s,T]|\Psi(\textbf{r})|}\;,
    \label{eq:HThermApprox}
\end{equation}

where $A$ and $B$ are best fit to match the exact value of the integral, see Ref.\,\cite{Kusch2026} for more details and also Ref.\,\cite{Yu2025Thermally} for a similar approach with an alternative form. To account for the dependences of $A$ and $B$ on $a_s$ and $T$, we rely on a functional form that we extract for $\edd<1$ (mean-field stable regime) and use as such with $\edd\geq 1$, see Ref.\,\cite{Kusch2026} for more details. 

\section{Ground-state simulations} \label{sec:simulations}

Ground states are calculated by solving Eq.\,\eqref{eq:TeGPE} in imaginary time using a split-step Fourier method. We implement grid sizes of $L_{x,y,z}= (48, 48, 24)\,\mathrm{\mu m}$ with grid spacings of $\Delta L_{x,y,z}= (0.1882, 0.1882, 0.3810)\,\mathrm{\mu m}$ for a trap of frequencies $\nu_{x',y',z}= (39, 20, 97)$\,Hz in all simulations, matching the experimental values. 

The dipolar interactions (second term of Eq.\,\eqref{eq:Uint}) are efficiently evaluated in Fourier space due to the convolution theorem. In order to avoid artificial self-interaction of alias Fourier copies of the system, a finite-range cutoff of half the grid widths $(L_{x,y,z}/2)$ is implemented, and it is ensured that the simulated densities are negligible within $L_{x,y,z}/4$ of the edge of the grids.

For the simulations presented in Fig.\,\ref{fig:protocol}(b), we set $T=60$\,nK and $N=1.5\times 10^5$, matching the experimental temperature and total atom number within error bars. We note that $N$ is matched to the total atom number and not only to the condensed atom number. 
Setting the TeGPE particle number to match only the observed condensed number of atoms yields states that have fewer droplet/stripe structures than observed in experiments. This effect has also been seen in other works on dipolar supersolids using the eGPE~\cite{Bland2022TwoDimensional,Casotti2024Observation,He2025Observation}, and has been attributed to the fact that the condensed number is extracted from TOF images which do not probe the full macroscopic wavefunction but only the fraction located within the droplets or stripes~\cite{Norcia2021Twodimensional}. 

Our choice yields simulated states whose characteristic sizes and structure numbers match more closely those seen in the experiment. Furthermore, the number of atoms located within the stripes or droplets in these simulated states is found to roughly match the experimental condensed atom numbers.

Fig.\,\ref{fig:protocol}~(b) in the main text shows the phase diagram as a function of $a_s$ and $\alpha$. To account for the fact that integrated densities can hide the three-dimensional nature of overlapping density structures introduced by the tilt, we consider a cut of the density through the plane $z=0$ to reliably distinguish between tilted droplets and stripes. 

In order to classify the ground states, we compute the power spectrum of the density cuts at $z=0$. We extract the smallest relative angular difference $\theta_\mathrm{min}$ between peaks in the power spectrum and classify it as triangular (stripe) if $\theta_\mathrm{min} \leq (>) \,90^\circ$.

\section{Imaging Sequence}\label{sec:imaging}

Both in-situ and TOF images presented in the main text are acquired by resonant absorption imaging on the 421\,nm Dy transition of linewidth $\Gamma =2\pi \times 32.2\,$MHz and saturation intensity $I_{\rm sat}= 56 \textrm{mW}/\textrm{cm}^2$. 
Imaging is performed along the $z$ axis using a microscope objective providing a theoretical resolution of $0.48\,\um$ at 421\,nm. The resolution has been tested via noise analysis on a 2D thermal gas, yielding $0.89\,\um$. The cloud is imaged on a qCMOS camera (Orca Quest) through an additional lens ($f=750\,\mathrm{mm}$), providing an overall magnification of $M=24(1)$. The magnification is calibrated by comparing the horizontal displacement of a thermal cloud, induced by an in-plane magnetic gradient of $6.7\,$G/cm applied for 17\,ms, as measured with vertical and horizontal imaging systems. The horizontal-imaging magnification is independently calibrated from a time series of a thermal cloud falling under gravity.
 
The imaging beam has a waist of $320\,\mathrm{\mu m}$ at the position of the atoms and power up to 3.3\,mW.  For in-situ imaging,  the saturation parameter $s=I/I_{\rm sat}$ is $s\approx36$, and an imaging pulse of $\tau = 5\,\mathrm{\mu s}$ is used. For TOF images, $s=1.2$ and $\tau = 25\,\mathrm{\mu s}$. During an imaging sequence, three frames are taken: one with the light and the atoms, $C_{\rm in}$, one with only the light (the atoms having been removed through free fall), $C_{\rm out}$, and one dark background image without light, $C_{\rm dark}$. We then calculate the optical density (OD) profile via \cite{Reinaudi2007lti,Horikoshi2017AbsorptionImg}: 
\begin{eqnarray} 
    OD=-\log \left( \frac{C_{\rm out}-C_{\rm dark}}{C_{\rm in}-C_{\rm dark}} \right)
    +\frac{C_{\rm in}-C_{\rm out}}{\chi_{sat}\cdot \tau}.
    \label{eq:OD}
\end{eqnarray}
The first term on the right-hand side corresponds to the low-intensity behavior (Beer-Lambert law). The second term accounts for saturation and becomes relevant for $s\gtrsim 1$.  Here, $\chi_{sat}\cdot \tau$ denotes the number of detected photons per pixel for $s=1$ and pulse length $\tau$. To calibrate $\chi_{sat}$ for our imaging system, we take a set of images of a thermal gas using varying $s$ and $\tau$, and minimize the variations of the OD extracted via Eq.\,\eqref{eq:OD}. This yields $\chi_{sat}= 1400$ counts$/\us$.
 
We note that the in-situ images of Fig.\,\ref{fig:geom} display negative OD values, in particular in between the high-density regions of the modulated states. We attribute these features to optical aberrations occurring because the axial extent of the states is larger than the depth of focus of our imaging system of $\lesssim 1\,\um$. From theoretical simulations, the axial extent of the states is estimated to be up to $2.2\,\um$. Other effects, such as collective scattering, might also contribute~\cite{Parit2025unveiling}, which warrants further investigation.

\section{Analysis of the in-situ images} \label{sec:insituanalysis}
 
Each in-situ image is rotated to align with the magnetic-field axes and zero-padded to increase the power-spectrum sampling. This results in $n(x=j p,y=l p)$ with $j \in [1,N_x], l \in [1,N_y]$ being the indices of the padded grid, and $p= 0.19\,\um$ the effective pixel size at the atom position. The power spectrum is computed via $S(k_x,k_y) = \left| \mathcal{F}\{n(x,y) \} \right|^{2}$ with wavenumbers $(k_x,k_y) = \frac{2\pi}{N_x p}(j,l)$, where $j$, $l$ are the indices of the Fourier grid.
 
We isolate the band $k_\rho \in [k_i = 1.3$\,$\mu\mathrm{m}^{-1}, k_f = 2.7$\,$\mu\mathrm{m}^{-1}]$ containing the finite-$\vk$ peaks. We identify the local maxima in this band by morphological dilation: a pixel is retained as a peak if and only if it equals the maximum of its local neighborhood inside the band. 

We restrict the subsequent analysis to a fixed angular window $\theta \in [30^\circ, 150^\circ]$. Distinct peaks are retained by imposing a minimum separation in $k$ of $\approx0.6$\,$\mu\mathrm{m}^{-1}$ and a minimum angular separation of $6^\circ$ before ranking them by amplitude. 

To delineate the boundary between the unmodulated and modulated phases, we compare the strongest identified finite-$\vk$ peak (power $S_1$) with the $\vk=0$ peak (power $S_0$). Accordingly, we define the ratio $R_{0} = S_1/S_0$. 

\begin{figure}[hb!]
     \centering
     \includegraphics[width=1.0\linewidth]{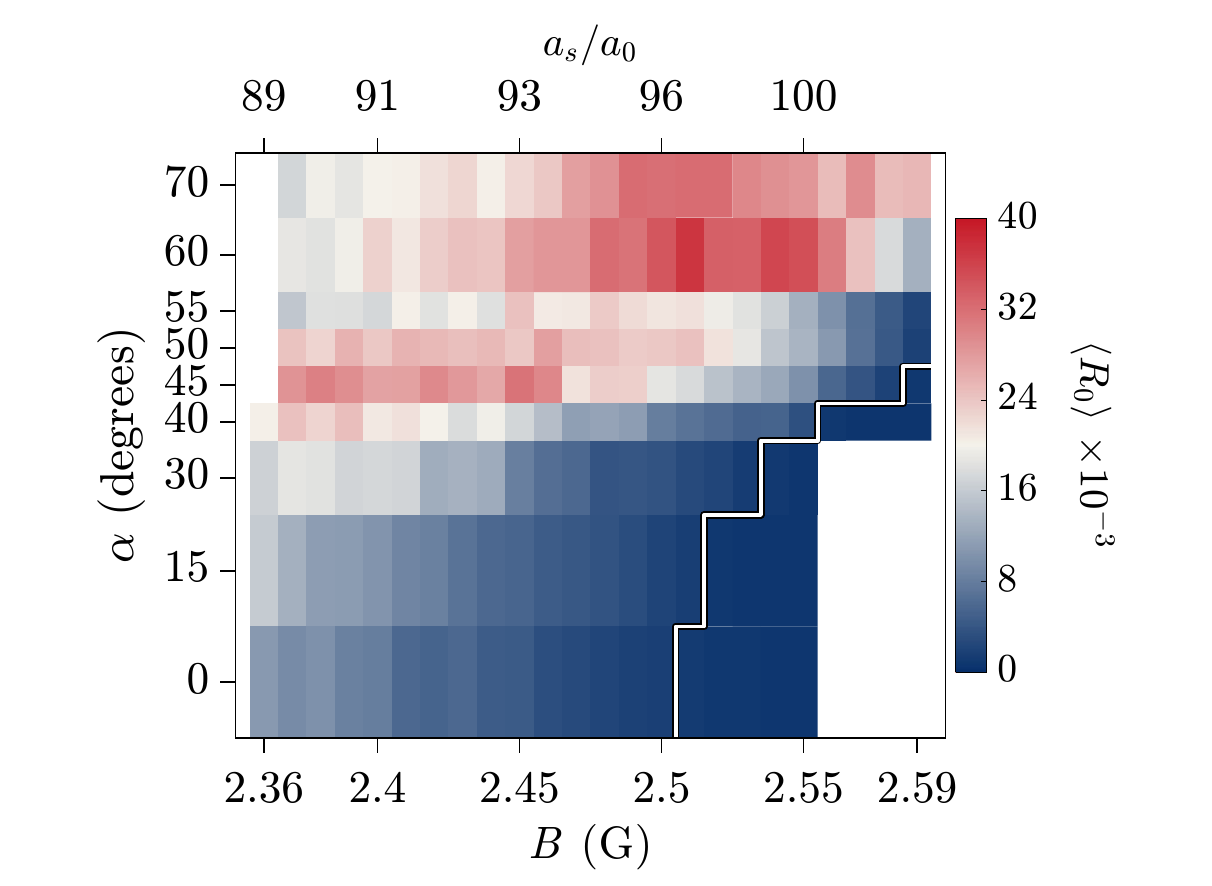}
     \caption{Mean value of $R_0$ as a function of the final $B$ and $\alpha$, with the solid white line indicating the boundary above which it grows beyond $1.2\,\permille$.}
     \label{fig:power_dc_ratio}
 \end{figure}

Figure~\ref{fig:power_dc_ratio} shows the mean of $R_{0}$ for the dataset of Fig.\,\ref{fig:opstatistics}. Starting from small $\alpha$ and large $B$, either decreasing $B$ or increasing $\alpha$ increases $R_{0}$. This indicates the onset of the density-modulated phase, which we identify by $R_{0}$ exceeding $1.2\,\permille$. The onset shifts to larger $B$ upon increasing $\alpha$ as theoretically expected, see Fig.\,\ref{fig:protocol}~(b). 

At large $B$ and small $\alpha$, corresponding to the unmodulated phase, the states are not entirely devoid of structure, as hinted by the small but non-zero $R_{0}$. We attribute these low-amplitude finite-$\vk$ peaks to the thermal population of roton-like excitations in our finite-temperature unmodulated superfluids ~\cite{Chomaz2018Observation,Schmidt2021Roton,Hertkorn2021Density}.

 \begin{figure}[ht!]
     \centering
     \includegraphics[width=1.0\linewidth]{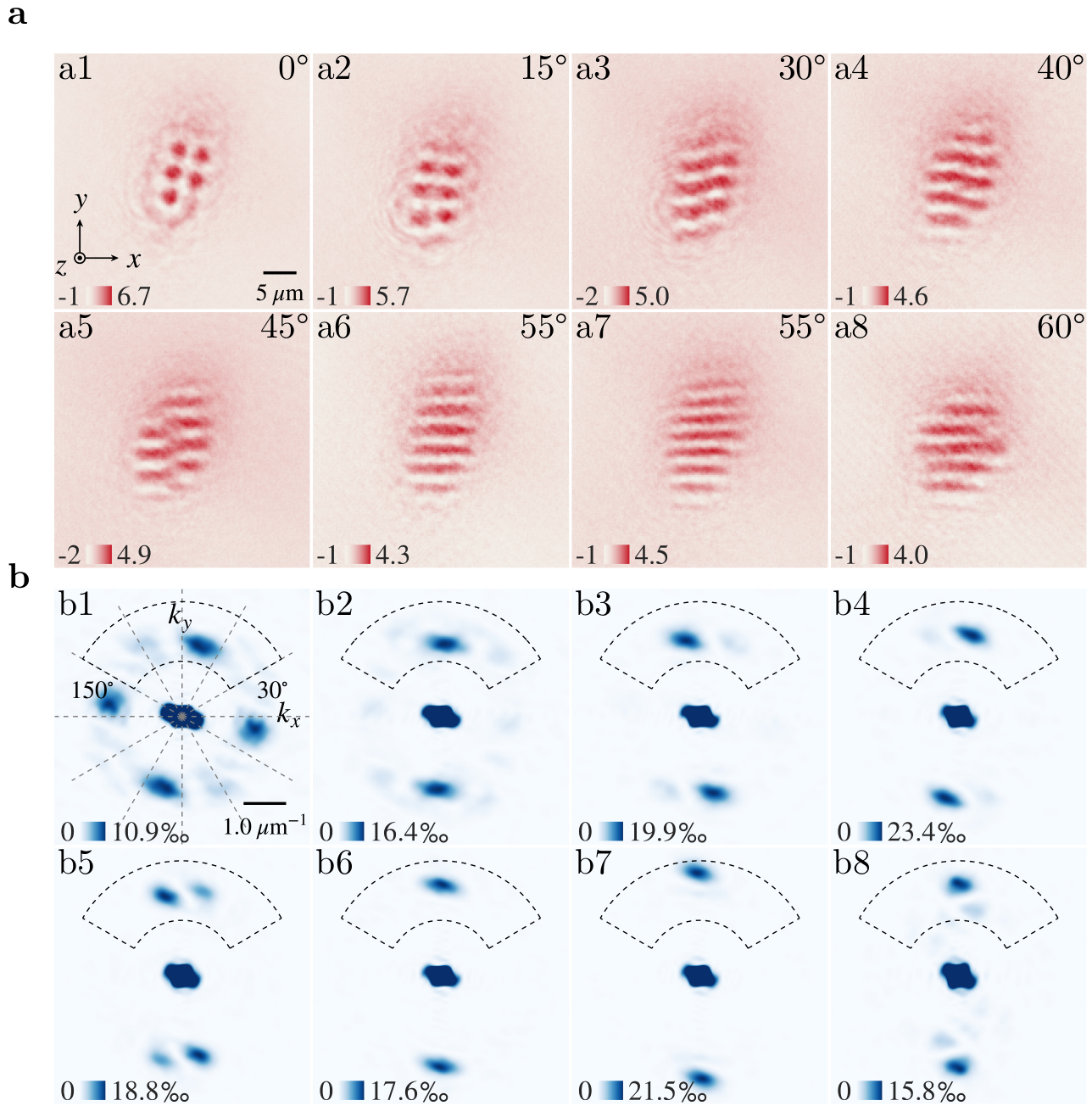}
     \caption{Selected states exhibiting distinct arrangements compared to Fig.\,\ref{fig:geom}(a5-8) at $B = 2.39\,\mathrm{G}$ and $\alpha=0^\circ$ (a1,b1), $15^\circ$ (a2,b2), $30^\circ$ (a3,b3), $40^\circ$ (a4,b4), $45^\circ$ (a5,b5), $55^\circ$ (a6-7,b6-7), $60^\circ$ (a8,b8). Both the in-situ density $n(x,y)$ (a1-8) and the corresponding power spectrum $S(\vk)$ (b1-8) are shown.}
     \label{fig:otherex}
 \end{figure}

In the modulated phase, $R_{0}$ varies with $\alpha$ and $B$, indicating changes in the density-modulation structure.  Decreasing $B$ at fixed $\alpha$ initially increases $R_{0}$, reflecting an increase in the contrast of the modulation. For large $\alpha$, this trend reverses when decreasing $B$ further, which relates to apparent excitations of the stripe structures. Changes of $R_{0}$ also occur upon increasing $\alpha$. These changes do not directly relate to the modulation contrast but are also influenced by the stretching of the droplets with $\alpha$ and the rearrangement from triangular arrays to stripes.

\begin{figure*}[ht!] 
    \centering
    \includegraphics[width=1.0\linewidth]{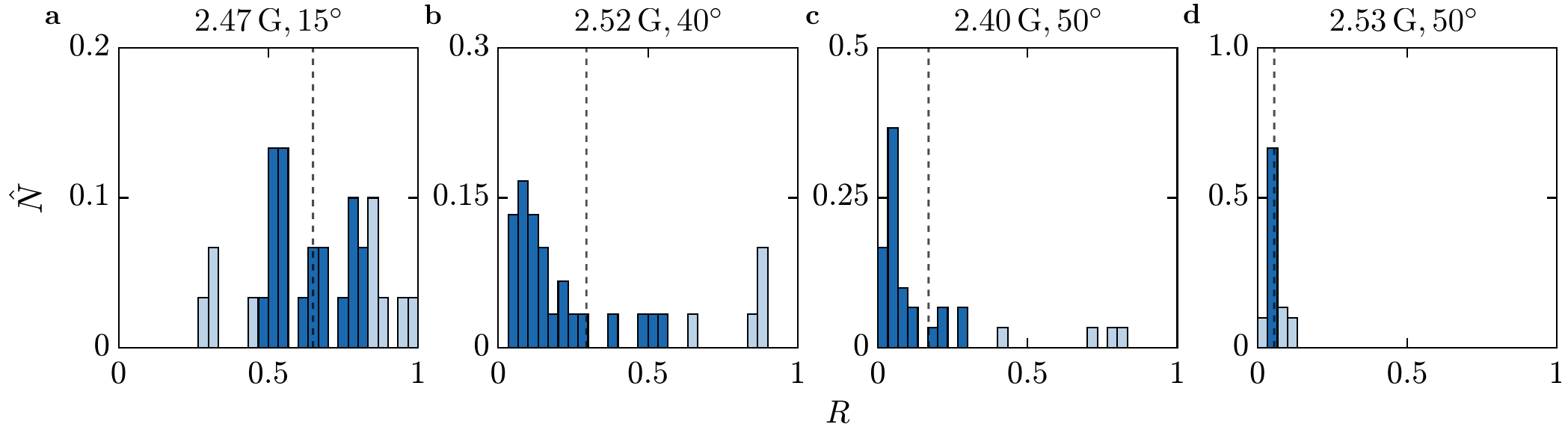}
     \caption{Experimental normalized distributions of the values of $R$ for 30 shots at $[\alpha, B, a_s]$ = $[15^\circ, 2.47\,\mathrm{G}, 94a_0]$ in the triangular phase (a), $[40^\circ, 2.52\,\mathrm{G}, 97a_0]$ (b) and  $[50^\circ,2.40\,\mathrm{G}, 91a_0]$ (c) both in the fluctuating regime, and $[50^\circ,2.53\,\mathrm{G}, 98a_0]$ in the stripe phase (d). The vertical dashed lines show the mean value $\kappa_1(R)$, and the dark blue bins fall into the $1\sigma=\sqrt{\kappa_2(R)}$ region around the mean.}
     \label{fig:R_distribution_exp}
\end{figure*}

\begin{figure*}[ht!] 
    \centering
    \includegraphics[width=1.0\linewidth]{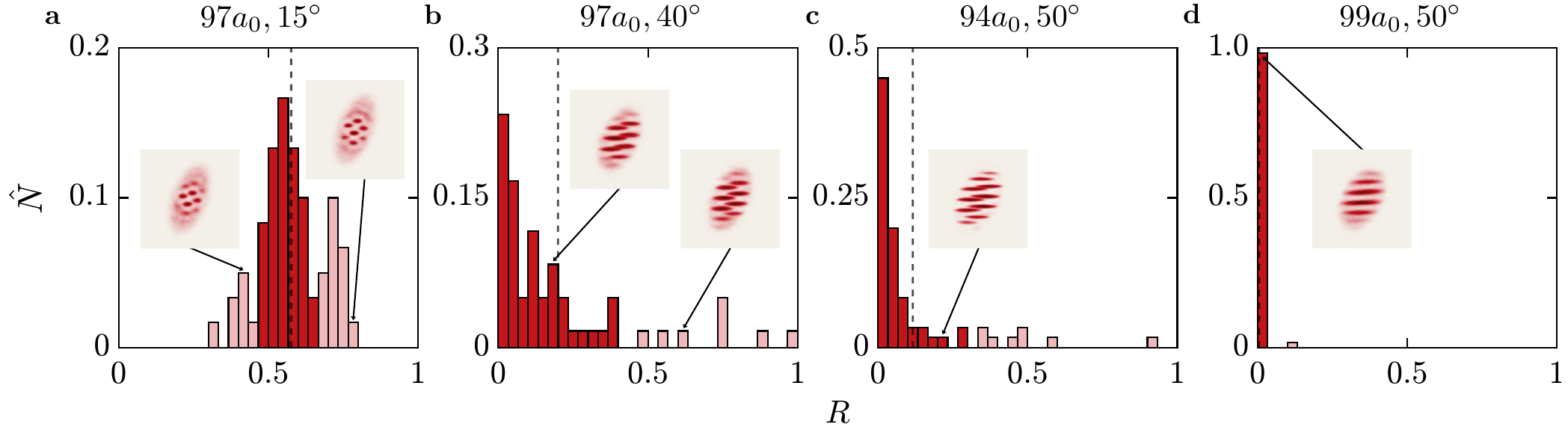}
     \caption{Theoretical normalized distributions of the values of $R$ from 60 repetitions of the TeGPE real-time simulations ramping to $(97a_0, 15^\circ), (97a_0, 40^\circ), (94a_0, 50^\circ)$ and $(99a_0, 50^\circ)$. The vertical dashed lines show the mean value $\kappa_1(R)$, and the dark red bins fall into the $1\sigma=1\sqrt{\kappa_2(R)}$ region around the mean. The insets show exemplary integrated densities at the indicated $R$ bin.}
     \label{fig:R_distribution_theo}
\end{figure*}

\section{Further examples of in-situ patterns and corresponding power spectra}\label{sec:furtherinsituexamples}
 
In the main text (Fig.\,\ref{fig:geom}), we show examples of in-situ images and the corresponding spectra for varying parameter values. Variations in the in-situ density patterns also occur from shot-to-shot when repeating the experiment under identical conditions. Fig.\,\ref{fig:otherex} shows measurements taken under the same conditions as those of Fig.\,\ref{fig:geom}(a5-a8) but exhibiting different structural arrangements. Here again the different structural arrangement translates to a different angular distribution of the finite-$\vk$ peaks in $S(k_\rho,\theta)$. In the case of droplet arrays,  typically several peaks appear in the upper ($k_y>0$) sectors, yet excitations of the regular triangular arrangement as seen for $\alpha = 0-15^\circ$ [Fig.\,\ref{fig:otherex}(a1,a2)] yield unequal amplitudes of these peaks [Fig.\,\ref{fig:otherex}(b2,b2)]. In the intermediate regime of $\alpha = 30-45^\circ$, the states are observed to fluctuate between staggered-triangular or wavy-stripe patterns, and sheared-array or straight stripe ones [Fig.\,\ref{fig:otherex}(a3,a4,a5)].  Several peaks, shifted in $\theta$, and of unequal amplitudes are observed in the upper ($k_y>0$) sectors [Fig.\,\ref{fig:otherex}(b3,b4,b5)].
Excitations of the stripe states for $\alpha > 45^\circ$  [Fig.\,\ref{fig:otherex}(a6,a7,a8)] might also lead to the emergence of additional peaks, of low amplitude, shifted mostly in $k_\rho$ [Fig.\,\ref{fig:otherex}(b8)], indicating a different periodicity of the modulation [Fig.\,\ref{fig:otherex}(b6,b7)].

\section{Examples of experimental distributions of the structural order parameter values}\label{sec:experimentalRDistributions}
 
In the main text (Fig.\,\ref{fig:opstatistics}), we examine the statistics of the order parameter over the parameter space by computing its first four cumulants. This allows us to identify three regimes, that of the triangular phase at small $\alpha$, the stripe phase at large $\alpha$, and a fluctuating intermediate regime indicating critical behavior. Figure \ref{fig:R_distribution_exp} shows the full distribution of the $R$ values over 30 experimental shots for four parameter sets covering these three regimes. 

\section{Examples of distributions of the structural order parameter values from simulation}\label{sec:theoryRDistributions}
 
We simulate the experimental sequence by evolving Eq.\,\eqref{eq:TeGPE} in real time in steps of $5\, \mu\mathrm{s}$. We initialize the wavefunction to the unmodulated ground state at $\alpha = 0^\circ$, $a_s = 100a_0$, and add random noise consisting of harmonic oscillator modes with amplitudes sampled from a Bose-Einstein distribution at $T=60\mathrm{nK}$. We let this initial state thermalize for $20 \, \mathrm{ms}$. We then ramp $\alpha$ and $a_s$ for $100\, \mathrm{ms}$ and hold for 30\,ms, following the experimental protocol [Fig.\,\ref{fig:protocol}(a)]. We record the $R$ value extracted from the power spectra of the integrated density profiles of the final state, as in experiment (Fig.\,\ref{fig:geom}). We show in Fig.\,\ref{fig:R_distribution_theo} the distribution of the $R$ values over 60 repetitions of the simulations, using independent noise. Similar to Fig.\,\ref{fig:R_distribution_exp}, we use four parameter sets covering triangular (a), fluctuating (b,c), and stripe (d) regimes. As in the experiment, the mean of $R$ is high in the triangular, low in the stripe, and intermediate in the fluctuating regime, with the largest variance found in this intermediate case.

\section{Sequence for probing phase coherence}\label{sec:sequence}

To probe phase coherence, the gas is imaged along $z$ after TOF. The imaging system is focused on the in-situ cloud position and has a limited depth of focus (App.\,\ref{sec:imaging}). To take TOF images, we do not realign the imaging path but bring the cloud back to the in-situ position using magnetic gradients. The TOF is split into two steps. First, we let the cloud fall and expand for 10\,ms while keeping the magnetic-field configuration as at the end of the sequence. Second, we apply a magnetic gradient for 40\,ms. The gradient and offset values during this second stage are chosen to return the cloud to its in-situ position at the end. This two-step scheme maintains the interactions at their initial strength during the early expansion, and delays the application of the gradient until interaction effects are substantially reduced. Interaction effects during the first stage are significant in both experiment and simulations. Interaction effects in the second step, and changes induced by the modified field configuration, cannot be eliminated completely but are minimized by the late switch time. Increasing the duration of the first step beyond 10\,ms requires larger gradients in the second step, leading to increased thermal effects in the coils and visible fluctuations in the cloud position.
 
\section{Analysis of phase coherence} \label{sec:coherence}

\begin{figure}[ht!]
     \centering
     \includegraphics[width=1.0\linewidth]{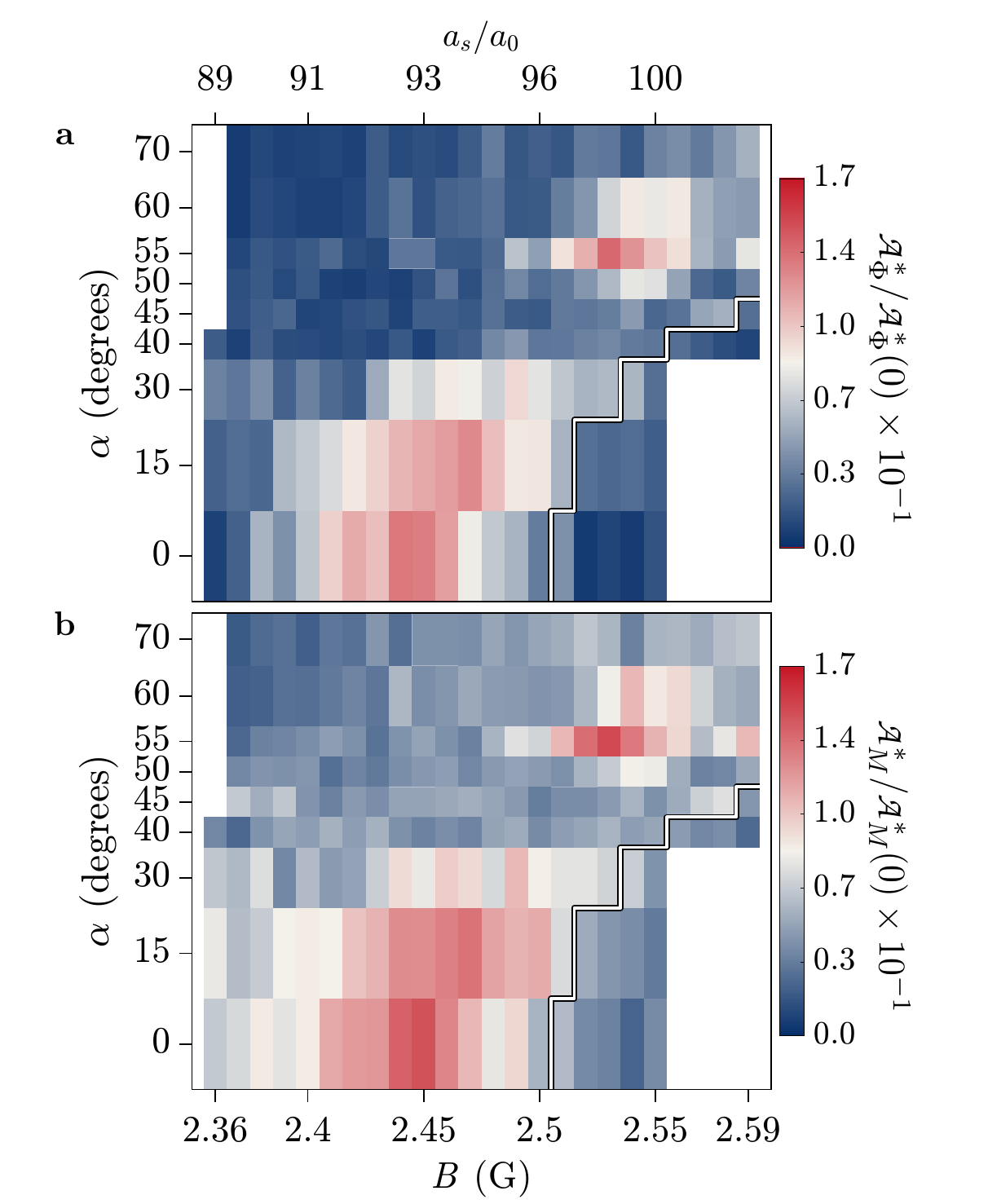}
     \caption{Amplitudes (a) $\mathcal{A}_{\Phi}^*$ and (b) $\mathcal{A}_{M}^*$ as a function of $B$ and $\alpha$. The values are normalized by the values of the profiles  $\mathcal{A}_{M}$ and $\mathcal{A}_{\Phi}$ at $y=0$, $\mathcal{A}_{M}(0)$ and $\mathcal{A}_{\Phi}(0)$, respectively. The solid white line highlights the boundary between modulated and unmodulated states as extracted from the power-spectrum analysis.}
     \label{fig:phasecoherence2}
 \end{figure}
 
To mitigate shot-to-shot center-of-mass fluctuations, we calculate the center of mass of each OD distribution and zero-pad the images to center them. For $\alpha>0$, the orientation of the structures is mainly set by the in-plane magnetic-field component. For $\alpha=0$, the triangular structures instead fluctuate in orientation. We compensate for these fluctuations by rotating the $\alpha=0$ images. We determine the angle of the strongest peak in the 2D Fourier transform of each TOF image and rotate the original image accordingly. The TOF images (Fig.\,\ref{fig:phasecoherence}) are shown as a function of $k_x,k_y$ obtained from the free-expansion formula,
$(k_x,k_y) = m(x,y)/\hbar t_{\mathrm{TOF}}$ where $m$ is the atomic mass, and $t_{\mathrm{TOF}} = 50\, \mathrm{ms}$. The aligned individual images [Fig~\ref{fig:phasecoherence}(a)], are then averaged [Fig~\ref{fig:phasecoherence}(b)] or used to calculate the Fourier-transformed integrated profiles $\mathcal{A}^{(i)}(y)$ and their averages $\mathcal{A}_{M}(y)$ and $\mathcal{A}_{\Phi}(y)$  [Fig~\ref{fig:phasecoherence}(c)].

Phase coherence is probed at a length $y^*$, namely $\mathcal{A}^*_{M}=\mathcal{A}_{M}(y^*)$ and $\mathcal{A}^*_{\Phi}=\mathcal{A}_{\Phi}(y^*)$. To determine $y^*$, we first extract a characteristic value $y^*_0(\alpha)$ for each angle $\alpha$ and define an interval centered on $y^*_0(\alpha)$ within which we search for $y^*$. 
We estimate  $y^*_0(\alpha)$ as the mean distance between central and side peaks in $\mathcal{A}^{(i)}(y)$ over all shots $i$ at a single magnetic field value in the phase-coherent regime. The extracted $y^*_0(\alpha)$ is used for all values of $B$ as we observe that the periodicity of the density modulation varies weakly with $B$. Using a common $y^*_0(\alpha)$ for all $B$ also avoids ambiguities in the unmodulated and phase-incoherent regimes. The length $y^*$ is then found within $\approx \pm 1.2 \: \mathrm{\mu m}$ of $y^*_0(\alpha)$ for each value of $B$. 

Figure~\ref{fig:phasecoherence}(d) discusses $\mathcal{A}^*_{\Phi}/\mathcal{A}^*_{M}$ as a measure of phase coherence. The individual amplitudes $\mathcal{A}^*_{M}$ and $\mathcal{A}^*_{\Phi}$ also provide relevant information, e.g. on modulation strength, density-structure fluctuations, and role of noise. Figures~\ref{fig:phasecoherence2} (a) and (b) show $\mathcal{A}^*_{M}$ and $\mathcal{A}^*_{\Phi}$  for the same dataset as Fig.\,\ref{fig:phasecoherence}(d). We normalize both quantities by the amplitudes at $y=0$, $\mathcal{A}_{M}(0)$ and $\mathcal{A}_{\Phi}(0)$, respectively, to partly account for atom-number variations (App.\,\ref{sec:atomlosses}).

 \begin{figure*}[ht!]  
     \centering
     \includegraphics[width=1.0\linewidth]{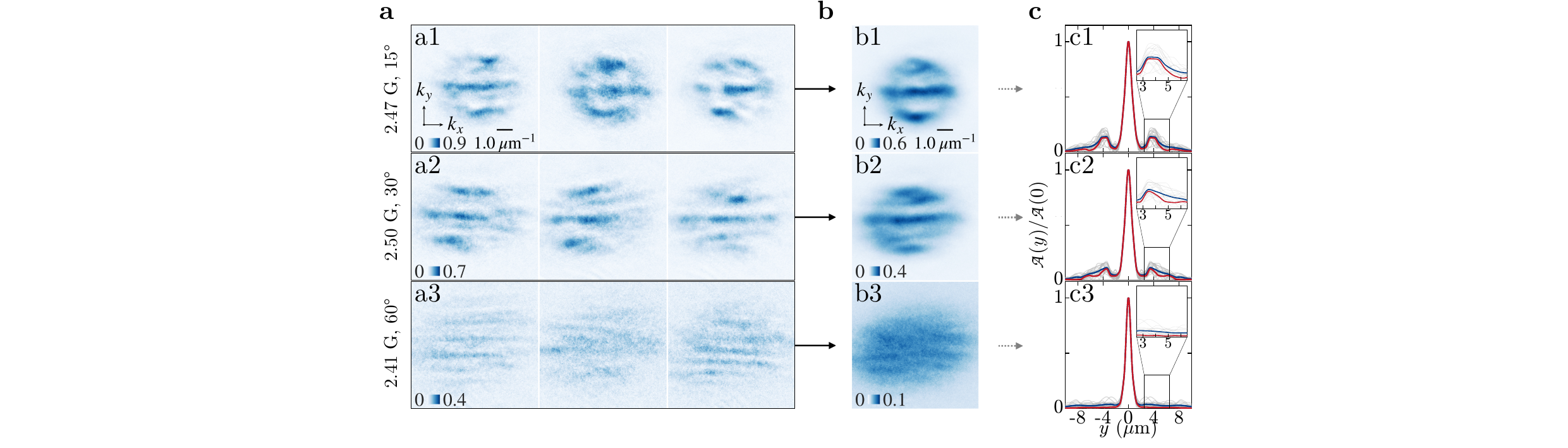}
     \caption{(a) More examples of single TOF density profiles, with (b) corresponding averaged images over 20 shots, shown as a function of the in-plane momentum components, $k_x$, $k_y$. (c) Corresponding 1D Fourier transforms of the individual density profiles integrated along $k_x$ (grey lines) together with $\mathcal{A}_\phi(y)$ (red line), $\mathcal{A}_M(y)$ (blue line). The inset in (c) enlarges the range around the $y>0$ side peak.  Same as Fig.\,\ref{fig:phasecoherence}(a,b,c) but now for: $B = 2.47\,\mathrm{G}$, $\alpha = 15^\circ$ (a1,b1,c1), $B = 2.50\,\mathrm{G}$, $\alpha = 30^\circ$ (a2,b2,c2), $B = 2.41\,\mathrm{G}$, $\alpha = 60^\circ$ (a4,b4,c4).}
     \label{fig:otherpcex}
 \end{figure*}
 
For decreasing $\as$ just below the unmodulated-to-modulated phase transition (white line), $\mathcal{A}^*_{M}$ and $\mathcal{A}^*_{\Phi}$ show increasing values, both at small $\alpha \leq 30^\circ$ (triangular arrays) and large $\alpha>50^\circ$ (stripes). This indicates the emergence of modulated states of increasing density contrast.  At intermediate $\alpha= 40-45^\circ$, both amplitudes remain low for all $a_s$, which we attribute to fluctuations of the density structures in the critical region. There, noise has a strong impact on the ratio $\mathcal{A}^*_{\Phi}/\mathcal{A}^*_{M}$. The amplitudes $\mathcal{A}^*_{M}$ and $\mathcal{A}^*_{\Phi}$ also decrease as the critical region is approached by increasing (decreasing) $\alpha$ for triangular (resp. stripes) arrangements, indicating progressively stronger fluctuations. At $\alpha=70^\circ$, both amplitudes remain low. The stronger atom losses observed at this angle (App.\,\ref{sec:atomlosses}) likely contribute to this reduced signal despite the normalization by $\mathcal{A}_{M}(0)$ and $\mathcal{A}_{\Phi}(0)$. For all $\alpha$, a further decrease of $\as$ strongly suppresses $\mathcal{A}^*_{\Phi}$ and also reduces $\mathcal{A}^*_{M}$, a behavior expected from the loss of global phase coherence~\cite{Hadzibabic2004Interference,Ilzhofer2021pci}.

\section{Further examples of phase coherence analysis}\label{sec:furtherphasecoherence}
 
In the main text [Fig.\,\ref{fig:phasecoherence}(a,b,c)], we show examples of our TOF analysis for four different parameters, corresponding to the four distinct regimes we identify, namely  phase-coherent triangular arrays, phase-coherent stripes, state in the critical regime, phase-incoherent states. Fig.\,\ref{fig:otherpcex} provides complementary examples at different tilt angles and scattering lengths, illustrating the cases of phase-coherent triangular arrays with dipoles at moderate tilt (a), and at larger tilt approaching the fluctuating region (b), and of phase-incoherent stripes (c).

\section{Ramp sequence and atom losses} \label{sec:atomlosses}

\begin{figure}[hb!]
     \centering
     \includegraphics[width=1.0\linewidth]{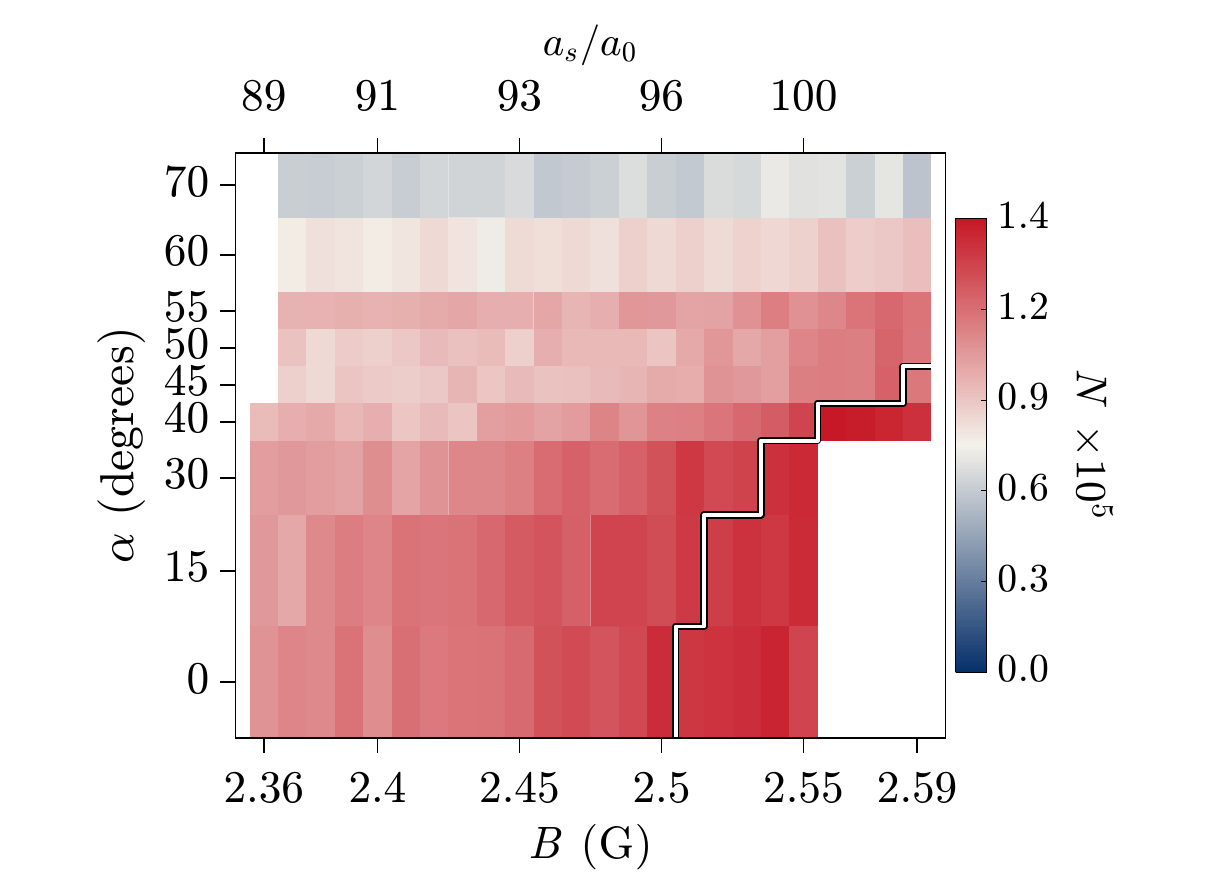}
     \caption{Total atom number measured at the end of the experimental sequence as a function of the final $B$ and $\alpha$. The solid white line highlights the boundary between modulated and unmodulated states as extracted from the power-spectrum analysis.}
     \label{fig:atnumbers}
 \end{figure}

After creating our initial cold samples, we simultaneously ramp $B$ and $\alpha$ in $\tr = 100\,$ms, and then hold the gas in the trap for $\tho = 30\,$ms, see Fig.\,\ref{fig:protocol}(a). As described in the main text, at the end of this sequence, different states are formed depending on $B$ and $\alpha$. As the states have both varying in-situ densities and scattering properties, losses occur differently as a function of $B$ and $\alpha$, resulting in different final atom numbers. We estimate the total atom numbers at the end of each ramp sequence by performing TOF measurements with 20 ms of free fall followed by horizontal imaging (App.\,\ref{sec:preparation}) and report them in Fig.\,\ref{fig:atnumbers}. We observe that the atom number decreases with decreasing magnetic field but the loss remains limited up to $\alpha \lesssim 55^\circ$, with $N\gtrsim 10^5$. For $\alpha =60-70^\circ$, atom losses become more severe and $N<8\times 10^4$ for all $B$ values.

\clearpage
\pagebreak
\bibliography{References}

\end{document}